# A multi-layered embedded intrusion detection framework for programmable logic controllers

Rishabh Das
*J. Warren McClure School of Emerging Communication Technologies*
Ohio University
Athens, USA
rishabh.das@ohio.edu

Aaron Werth
*Electrical and Computer Engineering*
*The University of Alabama in Huntsville*
Huntsville, USA
aww0001@uah.edu

Tommy Morris
*Electrical and Computer Engineering*
*The University of Alabama in Huntsville*
Huntsville, USA
tommy.morris@uah.edu

*Abstract*—Industrial control system (ICS) operations use trusted endpoints like human machine interfaces (HMIs) and workstations to relay commands to programmable logic controllers (PLCs). Because most PLCs lack layered defenses, compromise of a trusted endpoint can drive unsafe actuator commands and risk safety-critical operation. This research presents an embedded intrusion detection system that runs inside the controller and uses header-level telemetry to detect and respond to network attacks. The system combines a semi-supervised anomaly detector and a supervised attack classifier. We evaluate the approach on a midstream oil-terminal testbed using three datasets collected during tanker-truck loading. The anomaly detector achieves zero missed attacks, corresponding to 0.998 Matthews correlation. The supervised stage attains 97.37 percent hold-out accuracy and 97.03 percent external accuracy. The embedded design adds a median of 2,031 microseconds of end-to-end latency and does not impact PLC's cycle time. The proposed architecture provides a multi-layer embedded security that meets the real-time requirements of an industrial system.

## I. Introduction

Industrial control systems (ICS) manage safety-critical processes in sectors such as electric power, water utilities, chemical processing, and oil & gas terminals. The system operates through tightly coupled feedback loops, which use sensors to obtain the state of the physical process, controllers to operate the process, and actuators to enforce changes. Visibility in the process depends on the integrity of the measurement and the trustworthiness of the intermediary nodes, such as human machine interfaces (HMIs), historians, and engineering workstations. When an adversary compromises the measurement layer or a trusted node in the control path, the controller can enforce operations on falsified data. Such a scenario can push the plant towards unsafe operating states. Upstream devices like next-generation firewalls [1], perimeter IDS [2], SPAN port passive monitoring [3] provide little insight into attacks that originate inside the HMI or sensor path. As a result, the controller layer remains a blind spot for security assurance.

Attackers routinely target trusted nodes, including human–machine interfaces (HMIs) and engineering workstations (EWS) across industrial control system domains, including nuclear power plants [4], smart manufacturing [5], smart grids [6], and oil terminals [7]. Wang et al. [4] analyze real-world incidents in which adversaries gained internal footholds via compromised engineering workstations, bypassing perimeter defenses and causing network-level disruption in a nuclear power plant. Mern et al. [8] show a multi-stage attack, where an adversary compromises internal nodes, maintains persistence inside the industrial network, and triggers system-wide disruption on mission-critical segments. Similarly, Kelli et al. [6] enumerate eight attack scenarios, like replay, command injection, and denial-of-service, that manipulate Distributed Network Protocol version 3 (DNP3) endpoints to degrade smart-grid SCADA operations. Attacks through compromised internal nodes are hard to detect because they can manipulate the system without triggering standard security alerts [9].

Prior work addresses the PLC-side visibility gap by using techniques like (i) model-driven monitors of control behavior [10], (ii) firmware-level runtime monitors that track I/O activity, timing, and network access [11], (iii) memory baselining to detect program/data tampering [12], and (iv) temporal profilers of CPU usage and call sequences for resource-abuse exploits [13]. These solutions depend on vendor-specific payload parsers or reside outside the control loop, where commands from a compromised HMI still appear legitimate. We address this gap by embedding IDS inside the PLC to inspect packet-header telemetry at the controller. We evaluate the IDS in a simulated tanker truck loading operation on a to-scale model of a midstream oil terminal, built according to American Petroleum Institute (API) standards [7]. We validate the detection performance across six attack scenarios and quantify the impact on the PLC's network response and real-time control. The contributions of this research are as follows:-

- An embedded IDS that runs inside the PLC, inspects protocol-agnostic header telemetry, and mediates HMI/EWS traffic to the controller.
- A two-stage detection framework that couples a semi-supervised classifier with a supervised classifier and installs kernel-level iptables rules to contain volumetric floods.
- An empirical study on a to-scale midstream oil terminal operation demonstrating high detection accuracy and cross-scenario generalization under adversarial traffic.
- An evaluation of IDS overhead quantifying network latency and impact on PLC real-time control.

The rest of the paper proceeds as follows. Section II discusses related literature on embedded IDS in ICS, Section III outlines the architecture of the embedded IDS, Section IV discusses the testbed architecture and experimental setup, and Section V outlines the accuracy and impact of embedded IDS on the ICS operation. Section VI concludes the work and outlines future research areas.

## II. RELATED WORKS

The existing literature explores capabilities like logic integrity monitoring, behavior-based detection, and invariant monitoring inside PLCs. Allison et al. illustrate an embedded host-based IDS that uses native technologies in modern PLCs to monitor integrity and process control in realtime [14]. The approach detects control logic compromises and does not monitor dynamic network behavior. Garcia et al. demonstrate a verification framework that integrates into the PLC's scan cycle [15]. The framework uses embedded hypervisors to improve control-cycle awareness, but requires hardware reconfigurations, limiting deployment on legacy PLCs.

A number of studies focus on controlling logic integrity and monitoring invariants [16]. Yang et al. introduce a non-invasive system that uses control invariants between sensor data and PLC commands to detect intrusions. The system analyzes control logs and domain-specific logic information to detect attacks on the industrial process. Our approach can complement such invariant-based detection with network telemetry analysis. Similarly, Hailesellasie et al. outline a formal verification technique to detect deviations in industrial processes [17]. The method requires access to structured text and is less effective against network-based threats. Tan et al. demonstrate automatic rule generation from PLC code for Zeek-based IDS [18]. Such rules can integrate with our approach and add additional security layers in our multi-layer detection framework. Park et al. demonstrate an IDS that can monitor illegal PLC access and changes in process control commands [21]. Our work expands this approach by introducing network telemetry analysis within realtime constraints.

Ayub et al. focus on logic injection methods that exploit underused PLC features [19]. Their work motivates our focus on inline packet mediation at the controller, which inspects all network communication before it reaches the PLC logic engine. Similarly, our work can complement Huang et al.'s watermarking approach, which detects unauthorized modification of the control signal that drives the physical process [20].

What distinguishes our approach from the aforementioned works is that the present work uses a defense-in-depth strategy encompassing multiple layers instead of a single approach and focuses on a comprehensive analysis of network statistics. We also perform practical evaluations in a real ICS environment to ensure that the IDS meets the realtime constraints of the industrial environment.

## III. ARCHITECTURE OF THE EMBEDDED IDS

This section outlines the architecture of the embedded intrusion detection system. The embedded IDS operates as a multi-layer defense inside the programmable logic controller (PLC), mediating all communication between external clients (like HMI and EWS) and the PLC logic. The internal placement allows the IDS to detect threats from the compromised Level 2 node (using the Purdue model nomenclature [7]). Level 2 nodes, such as an engineering workstation or HMI, have network access to the PLC, and can inject control commands, eavesdrop on industrial communication, and disrupt PLC communications. The attacker can operate with the privileges of a legitimate system node and exploit the implicit trust and lack of internal segmentation to carry out attacks that traditional perimeter IDS or firewalls might not detect. Embedding the IDS at the PLC enables visibility at the edge, and the IDS can inspect all incoming network traffic on the PLC's normal listening port. After multi-layer inspection, only safe packet flows to the PLC.

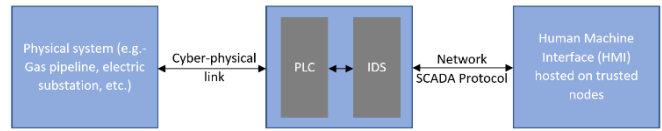

Fig. 1. Logical architecture of the Industrial Control System

The rest of the section discusses the five modules of the embedded IDS: the relay server module, data sensor, data preprocessor, analysis engine, and incident response system. Figure 1 illustrates the five logical components of an ICS with the IDS-enabled PLC. Figure 2 depicts the internal modules of the IDS.

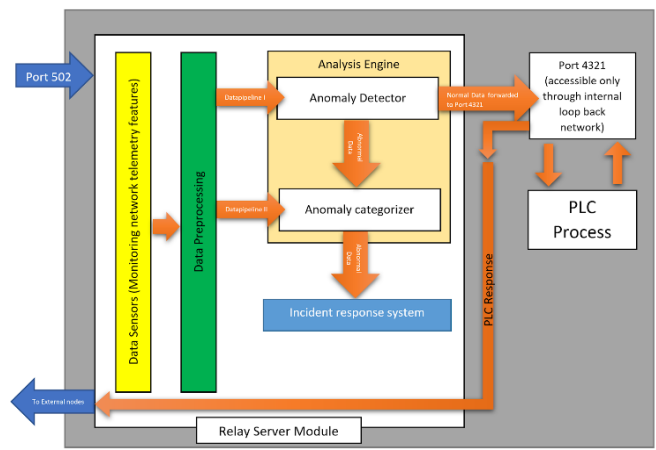

Fig. 2. Internal modules of the embedded IDS

### A. Relay Server Module

The relay server module is an embedded TCP proxy server application. It acts as an intermediary between the external nodes (HMI and trusted nodes) and the PLC control services. This research uses an open-source PLC application called OpenPLC [22]. OpenPLC allows users to configure the listening port as a runtime command-line argument. The relay server uses this feature to reassign OpenPLC's listening port from 502 (default for MODBUS) to port 4321. After reassignment, the relay server listens on port 502 and uses an iptables rule to block direct external access to the PLC's new listening port 4321.

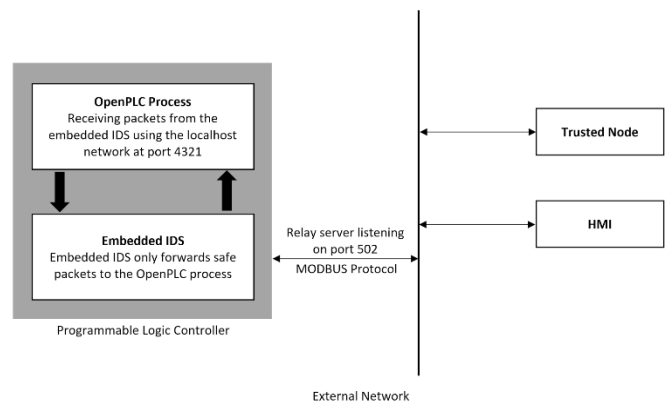

Fig. 3. External and internal network of the PLC with the embedded IDS

In normal operation, HMIs and other clients connect to the relay and send Modbus requests. The relay then intercepts and evaluates each packet before forwarding it to the PLC's actual listener on 4321 through an internal loopback network. During the evaluation, the relay server passes the network packet through the data sensor, data preprocessing, analysis engine, and incident response system. Figure 3 shows the internal and external network of the PLC.

*B. Data Sensor*

The data sensor module monitors incoming packets and extracts network telemetry features from each connection. In essence, the module acts as a real-time network traffic logger inside the PLC, and quantifies the behavior of each external peer device (HMI or trusted node) communicating with the PLC. For each packet, it captures L2–L4 header fields (lengths, ports, MAC/IP tuples) and high-resolution arrival timestamps, but does not parse application payloads. The header-only design keeps processing lightweight, and remains protocol-agnostic across Modbus/TCP, DNP3/TCP, and IEC-104, and even with encrypted PDUs (like TLS and SSL tunnels) [11]. In total, the module computes 14 network-telemetry features including three per-packet and eleven per-connection.

1) Per-packet-metrics are stateless snapshots from individual network packets. The data sensor computes these features as each packet arrives. The data sensor computes three per-packet-metrics for each incoming packet $p_{i,k}$ (the $k$-th packet received from peer $i$)

i. *Number of peers ($N_{\text{peers}}(k)$):* the count of unique source MAC addresses seen communicating with the PLC. A sudden jump can reveal a scan, rogue host, or bot amplification stage.

ii. *Packet size ($L_{i,k}$):* the total size of the network packet in bytes. $L_{i,k}$ provides a volumetric measure for flowrate, variance and scaling metrics.

iii. *Protocol efficiency ($E_{i,k}$):* the ratio of TCP payload bytes to total packet bytes, capturing protocol overhead. The ratio measures the data content on the packet. Spikes toward 1.0 signal large file transfers, while dips toward 0 indicate keep-alive or scanning chatter.

$$E_{i,k} = \frac{\ell_{i,k}}{L_{i,k}} \in (0,1] \quad (1)$$

Where, $\ell_{i,k}$ are the bytes in the application payload and $L_{i,k}$ is the total size of the packet in bytes.

2) Per-connection-metrics are features from two or more packets within the same logical conversation or peer. The sensor maintains per-peer counters, sliding windows, and other state to derive these time-dependent features. The data sensor computes eleven per-connection metrics.

iv. *Mean flow ($F_i(k)$):* average number of packets per 60-second interval from a specific peer. The 60-second time windows help filter traffic bursts so that the IDS can categorize DoS floods and momentary jitter.

$$F_i(k) = \frac{N_i(k,60s)}{60} \ [\text{pkt s}^{-1}] \quad (2)$$

Where, $N_i(k, 60s)$ number of packets from peer i seen during the last 60 s.

v. *Inter-packet arrival time ($\Delta t_{i,k}$):* measures the frequency at which a connected peer node sends network packets to the PLC. It is the time gap between consecutive packets from a specific peer.

$$\Delta t_{i,k} = t_{i,k} - t_{i,k-1} \quad (3)$$

Where, $t_{i,k}$, and $t_{i,k-1}$ are arrival times of two consecutive packets from the same peer.

vi. *Moving average of packet size ($\mu_{i,k}$):* the average size of the network packet in bytes over a rolling window of 1000 network packets. On receiving a new sample, the window slides forward and discards the oldest sample.

$$\mu_{i,k} = \frac{1}{N_w}\sum_{p \in W_{i,k}} L_p \quad (4)$$

Where, $W_{i,k}$ is a window containing the most recent $N_w (= 1000)$ packets from peer $i$.

vii. *Moving variance of packet size ($\sigma^2_{i,k}$):* Packet size variance measures how far the network packet sizes are spread out from their average value. The data sensor uses a rolling window holding 1000 previous network packet sizes for computing the moving variance. Moving variance large swings can help identify normal cyclic traffic from command injection or file exfiltration bursts.

$$\sigma^2_{i,k} = \frac{1}{N_w}\sum_{p \in W_{i,k}} (L_p - \mu_{i,k})^2 \quad (5)$$

Where, $\mu_{i,k}$ is the moving average from metric 6.

viii. *Moving median of packet size ($med_{i,k}$):* Similar to the moving mean and moving variance, the data sensor calculates the moving median over a rolling window containing 1000 previous network packet sizes. Moving median values are not influenced by outliers in the data sequence and provide unskewed central values of network packet sizes.

$$med_{i,k} = \text{median}\{L_p \mid p \in W_{i,k}\} \quad (6)$$

Where, $W_{i,k}$ is a window containing the most recent $N_w (= 1000)$ packets from peer $i$.

ix. *Scaled packet size ($sL_{i,k}$):* current packet size normalized by the maximum packet size observed from that peer.

$$sL_{i,k} = \frac{L_{i,k}}{\max_{j \leq k} L_{i,j}} \quad (7)$$

x. *Scaled inter-packet arrival time ($s\Delta t_{i,k}$):* current inter-arrival time normalized by the maximum gap observed for that peer. Smaller values can flag unusually fast transmissions or flooding.

$$s\Delta t_{i,k} = \frac{\Delta t_{i,k}}{\max_{j \leq k} \Delta t_{i,j}} \quad (8)$$

xi. *Source port count ($P_i(k)$):* the number of unique TCP or UDP source ports used by each connected peer node over the last 60 second time window. Concurrent connections from compromised nodes can increase the source port count.

$$P_i(k) = |\{\text{sport}(p_{i,j}) \mid t_{i,j} \in [k - 60s, k]\}| \quad (9)$$

Where, $\text{sport}(p_{i,j})$ is the TCP/UDP source-port value.

xii. *Clients-per-MAC ($C_m(k)$):* the total size of the network packet in bytes. $L_{i,k}$ provides a volumetric measure for flowrate, variance and scaling metrics. This metric helps detects ARP spoofing or MAC hijacking when multiple IPs present the same MAC.

$$C_m(k) = |\{IP \mid MAC(IP) = m\}| \quad (10)$$

Where, m is a MAC address of a peer.

xiii. *Entropy of packet size ($H_{i,k}$):* Shannon entropy of the distribution of the peer's recent packet sizes of last 1000 packets, measuring the unpredictability or information content of packet size patterns. A high entropy indicates highly variable packet sizes, whereas lower entropy indicates a stable, repetitive traffic pattern. Sudden changes in this entropy can signify anomalies

$$H_{i,k} = -\sum_{b \in B} p_{i,k}(b) \log_2 p_{i,k}(b) \quad (11)$$

Where, $B$ is pre-defined size bins (set to 5 bins). $p_{i,k}(b)$ is the relative frequency of sizes falling into bin $b$ inside the current window.

xiv. *Kullback–Leibler divergence of current connectivity characteristics to baseline data ($D_{i,k}$):* the divergence between the current packet size or inter-arrival distribution (over the recent window) and a baseline "normal" distribution precomputed for that peer during a training phase. A large KL divergence means the current traffic statistical profile deviates significantly from historically normal behavior.

$$D_{i,k} = \sum_{b \in B} P_i^{\text{base}}(b) \log_2 \frac{P_i^{\text{base}}(b)}{p_{i,k}(b)} \quad (12)$$

Where, $P_i^{\text{base}}(b)$ is the bin frequency learned from long-term benign traces, and $p_{i,k}(b)$ is the current-window frequency

The feature set is grounded in prior network anomaly-detection research that targets distributional shifts in network telemetry [28–31]. The per-packet and the per-connection metrics capture rate, timing, size dispersion, and flow/port diversity, alongside entropy and KL-divergence to quantify deviations from a benign baseline. After calculating the network telemetry metrics, the data sensor forwards a tuple of information to the data preprocessing unit.

*C. Data Preprocessing*

The feature tuples from the data sensor feed into the Data Preprocessing Unit, which prepares the data for analysis by two parallel detection pipelines.

The data pipeline I produces a normalized full-feature vector for a semi-supervised anomaly detection algorithm. The pipeline first uses min-max normalization to rescale each feature onto a fixed, bounded range to equalize influence across the fourteen features. Then we compress the information of fourteen features into a two-dimensional vector space using principal components analysis (PCA). For the midstream oil terminal datasets discussed in Section IV, choosing two PCA components generates a cumulative variance of 0.9873 for dataset I. The high variance score demonstrates that two PCA vectors are sufficient for compressing the information provided by the network telemetry features. Transformation parameters like normalization bounds and PCA model (selected component count) are fit once on the static baseline datasets (described in sections IV–V) and frozen during online operation.

The data pipeline II provides a feature vector optimized for a supervised classifier algorithm. The pipeline starts by rescaling the values using min-max normalization and then prunes highly collinear attributes with a correlation coefficient greater than 0.9. The remaining features undergo recursive feature elimination to retain only the most discriminative predictors. In our deployment, this process reduces the input tuple to five features: mean flow rate, packet size, moving average of packet size, protocol efficiency, and source-port count, which captures volume, size dynamics, and connection usage patterns. Table II in section V discusses the rationale behind choosing the data transformation and empirically proves that the preprocessing improves classifier accuracy.

*D. Analysis Engine*

The analysis engine receives the transformed data pipelines and feeds them into the anomaly detector and the anomaly categorizer.

*Anomaly Detector:* The data pipeline I supply transformed tuple into a semi-supervised novelty detection algorithm that outputs an alert if incoming traffic deviates significantly from normal patterns. We use the Local Outlier Factor (LOF) algorithm because it is effective in detecting novel or rare attacks in network traffic [25]. LOF is a density-based method introduced by Breunig et al. that compares the local density of an observation to that of its neighbors [26]. Trained on benign baseline data, the model captures the density envelope of normal traffic. At runtime, the algorithm scores each observation from Pipeline I. Values close to 1 indicate density comparable to neighbors (normal), while substantially larger values indicate the observation resides in a sparser region (anomalous). The neighborhood size (k) governs sensitivity. For example, if k is set to 5, k-distance will be the distance of the point to the fifth closest neighbor. A higher value of k finds outliers against the entire data set, while a lower value distinguishes outliers from local clusters. We optimize the value of k to yield the lowest number of false negatives. Section V A discusses the tuning process in detail.

Observations classified as benign are relayed to the PLC process. Observations flagged as anomalous are forwarded to the anomaly categorizer, which uses the Pipeline II feature set to assign an attack label.

*Anomaly Categorizer:* For this classification task, we employ a Random Forest (RF) classifier, chosen for its resilience against imbalanced datasets and accuracy in intrusion detection applications [27]. The random forest classifier categorizes the anomaly into six classes: MITM based eaves dropping(EX-4), a man-in-the-middle scenarios that drops all traffic between the HMI and the PLC causing a denial of service (EX-1), sensor and actuator spoofing scenarios (EX-2, EX-3), a command injection scenario (EX-6), and a volumetric denial of service using a flooding tool called Hping3 (EX-7). The traffic monitoring module forwards the categorized attack type and the IP and MAC address of the suspected attacker to the incident response system.

*E. Incident response System*

The incident response system executes a pre-defined set of rules to defend the PLC against the identified category of cyber-attack. On detection of a volumetric attack like the DoS attack or a repeated command injection attack, the incident response system logs the event and executes a custom IP table rule. The IP table rule drops all incoming traffic from the attacker's IP and stops the volumetric flood of packets from

This work was supported at least in part by the National Science Foundation through Grants 1623657 and 1431484

the attacker's computer. For the MiTM attack, the IDS logs the IP of the possible attacker causing the anomalous network condition. A user can add rules to take a unique set of incident responses against identified attacks.

IV. TESTBED AND DATASETS

This research integrates the embedded intrusion detection system inside PLCs controlling a simulated mid-stream oil terminal [7]. The oil terminal uses twelve PLCs to control liquid cargo operations, adhering to the American Petroleum Institute (API) standards. The PLC utilizes the OpenPLC suite, which is an IEC 61131–3 compliant open-source industrial controller, for managing operations and runs on a Raspberry Pi 3 (1 GHz with 1 GB ram). Prior research on the midstream oil terminal testbed using OpenPLC as the controller device showed high-fidelity responses during cyber-attack scenarios and illustrated the impact of cyber-attacks on connected processes [7].

The threat model presumes that the malicious attacker has established a path through the defensive strategies positioned in Purdue levels 5, 4, and 3 [7], and gained limited or complete control over the trusted node at level 2 of the defense-in-depth architecture. In such a scenario, the attacker from the compromised node can monitor and control the core services of the control system at level 2 or below. Prior work uses Khan et al. 's threat modeling technique to identify attack scenarios relevant to the oil terminal architecture [24, 28]. This research simulates six attack scenarios from the threat model. The attack scenarios includes a eavesdropping scenarios (EX-4), a man-in-the-middle scenarios that drops all traffic between the HMI and the PLC causing a denial of service (EX-1), sensor and actuator spoofing scenarios (EX-2, EX-3), a command injection scenario (EX-6), and a volumetric denial of service using a flooding tool called Hping3 (EX-7) [11, 7]. The attacks are during a Tanker Truck (TT) loading operation in a midstream oil terminal.

We use an automation script to capture four datasets. Dataset I capture network telemetry during a normal TT loading operation, containing 138,352 tuples. The semi-supervised data pipeline uses these datasets to baseline the system's normal behavior. Datasets II and III capture normal and anomalous data during a Tanker Truck loading operation in a midstream oil terminal. Table I shows the distribution of samples in the datasets.

TABLE I. NUMBER OF NORMAL AND ATTACK SAMPLES IN DATASETS

| Sample Type | Data Label | Number of data samples | | |
|---|---|---|---|---|
| | | *Dataset I* | *Dataset II* | *Dataset III* |
| Normal Samples | Normal | 138,352 | 135352 | 46704 |
| Anomalous Samples | EX-1 | No Attack Samples | 3106 | 3228 |
| | EX-2 | | 4212 | 4226 |
| | EX-3 | | 4253 | 4267 |
| | EX-4 | | 4237 | 4234 |
| | EX-6 | | 9997 | 10432 |
| | EX-7 | | 3023 | 3139 |

V. RESULTS AND DISCUSSION

This section outlines several experiments to benchmark the accuracy of the semi supervised and supervised data pipelines, measures the embedded IDS's impact on the PLC's real-time performance, and evaluates the performance of the IDS in blocking volumetric denial of service attacks.

*A. Accuracy of semi-supervised LOF anomaly detector*

Pipeline I learns the normal baseline for the trusted node and HMI using Dataset I. Each feature is first rescaled to [0,1] via min–max normalization [11]. We project the normalized vectors onto a two-dimensional space with PCA. The first two principal components explain 98.73% of the variance in Dataset I, which is sufficient to preserve the telemetry information. We fit the LOF model on these 2-D embeddings to capture the local density of benign behavior. To tune the neighborhood size, $k_{LOF}$, we refit LOF for $k_{LOF} \in \{5, \ldots, 24\}$ and evaluate on Dataset III. The entire procedure is repeated 20 times with different splits, and we select the $k_{LOF}=10$ that yields the lowest false-negative rate.

We evaluate the LOF anomaly detector on Dataset II (28,828 anomalous and 135,352 normal samples). The model yields True Positive (TP) = 28,828, True Negative (TN) = 135,255, False Positive (FP) = 97, and False Negative (FN) = 0. This corresponds to accuracy 99.94%, precision 99.66%, recall 100.00%, specificity 99.93%, F1 99.83%, and Matthew's correlation coefficient (MCC) 0.998. The MCC integrates all four cells of the confusion matrix and is widely recommended for imbalanced datasets [23]. High MCC confirms robust performance with a 0.072% false-positive rate.

*B. Accuracy of supervised datapipeline II*

Pipeline II classifies the abnormal traffic into Attack categories using a supervised random-forest classifier (200 trees). This experiment repeats the performance evaluation for four configuration, each applying a different preprocessing pipeline before feeding into the classifier: (i) using unprocessed features, (ii) min-max normalizing the features to [0,1], (iii) normalization plus a correlation filter that prunes highly co-linear features, and (iv) using a combination of normalization, correlation filter, and recursive feature elimination (RFE). RFE iteratively refits the forest and drops the least informative feature until the model reaches the highest validation accuracy. The RFE process yields a compact five-feature decision set containing mean flow, packet size, moving mean of packet size, protocol efficiency, and number of source ports. For each configuration, we fit the model on 75% of Dataset II and validate on the remaining 25%. We test the generalization on an unseen dataset (Dataset III) and compute the accuracy for all four configurations.

TABLE II. A SUMMARY OF TRAINING AND TESTING ACCURACY OF THE RANDOM FOREST CLASSIFIER

| Preprocessing configuration | Training Accuracy (75% of Dataset II) | Testing Accuracy (Hold-out 25 % of Dataset II) | External Testing Accuracy (Dataset III) |
|---|---|---|---|
| All features | 0.94251 | 0.92236 | 0.90354 |
| Normalization | 0.94571 | 0.92244 | 0.90379 |
| Normalization + Correlation filter | 0.94567 | 0.93263 | 0.91385 |
| Normalization + Correlation filter + Selected feature using RFE | 0.98421 | 0.97367 | 0.97027 |

Table IV outlines accuracy for the random forest classifier under four preprocessing configurations. Column 1 provides the training accuracy on the 75% split of Dataset II, Column 2 shows hold-out accuracy on the remaining 25% of Dataset II, and Column 3 reports external accuracy on unseen Dataset III. Because random forests are largely scale-invariant, normalization yields only marginal changes. Adding a correlation filter improves the hold-out/external accuracies to 0.93263/0.91385, indicating that pruning redundant, highly collinear features reduces variance. The model achieves the best results after applying RFE. Accuracy rises to 0.98421 (train), 0.97367 (hold-out), and 0.97027 (external) gains of +5.1 and +6.7 percentage points over the baseline hold-out/external scores. The small gap between hold-out and external results ($\approx 0.003$) shows strong generalization.

*C. Effect of the embedded IDS on the SCADA system*

This section outlines two tests to measure the impact of the IDS on the network response time and the real-time performance of the PLC.

*a) Impact on response time of the PLC*: This experiment measures the network latency of the IDS during eight MODBUS operations, like read coil, read discrete input, read holding register, read input registers, write single coil, write single holding register, write multiple coils, and write multiple holding registers. In each MODBUS operation, the HMI initiates a request directed at the PLC. When the operation is complete, the PLC issues a response to the HMI. This experiment measures the latency across 500 request-response cycles for each MODBUS operation. We repeat the experiment with and without the IDS to track the response times.

Table III shows the network latency in microseconds. The high standard deviation values show the network latencies have outliers. Median is a more robust statistical measure against outliers and skewed data than the mean, and is usually the preferred measure of central tendency when the distribution is not symmetrical. The median values indicate that the IDS introduces an average overhead of 2031 microseconds between the PLC and the HMI. The spread of the latencies is also narrow, illustrating stable timing behavior across different MODBUS functions. Accounting for upper-tail measurements, the response times constitute only a small fraction of a 50-millisecond control deadline for the Oil & Gas plant, and the experiment does not see any deadline-violating outliers.

TABLE III. RESPONSE TIME OF THE PLC IN MICROSECONDS

| MODBUS Operation | Without IDS | | | With IDS | | |
|---|---|---|---|---|---|---|
| | *Mean* | *Std. Dev.* | *Median* | *Mean* | *Std. Dev.* | *Median* |
| Read Coil | 1370.32 | 2107.99 | 602.07 | 3151.54 | 2485.15 | 2134.32 |
| Read Discrete Input | 1012.35 | 2136.46 | 368.99 | 3626.39 | 2627.34 | 2488.09 |
| Read Holding register | 669.29 | 1375.93 | 353.58 | 3305.56 | 2449.79 | 2452.94 |
| Read Input Register | 747.35 | 1740.86 | 339.16 | 3136.09 | 2151.87 | 2429.43 |
| Write Single Coil | 646.62 | 1410.55 | 321.58 | 3183.82 | 2224.77 | 2424.29 |
| Write Single Holding Register | 642.74 | 1474.27 | 325.33 | 3124.65 | 2127.97 | 2431.01 |
| Write Multiple Coils | 585.72 | 1252.26 | 333.63 | 3067.57 | 2088.58 | 2417.58 |
| Write Multiple Holding Registers | 668.82 | 1572.72 | 316.25 | 3441.93 | 2569.72 | 2430.81 |

*b) Effect on real-time behavior of the PLC:* This experiment analyzes the effect of embedded IDS on the real-time performance of the PLC. The cycle time of the PLC is set to 50 milliseconds, and an embedded logger monitors the cycle time for 24 hours. We repeat the experiment twice to study the cycle time data with and without the embedded IDS. Data over 1728000 observations with no embedded IDS records has a mean cycle time of 50.04 milliseconds and a standard deviation of 0.0142. Repeating the experiment with the embedded IDS detects no shift in the mean or increase in jitter within the logger's resolution, as shown in Table IV. The IDS adds no measurable delay to the PLC's real-time operation.

TABLE IV. PLC CYCLE TIME IN MILLISECONDS

| | Average | Standard deviation |
|---|---|---|
| PLC without IDS | 50.04 | 0.0142 |
| PLC with IDS | 50.04 | 0.0142 |

*D. IDS response time during volumetric denial of service attack using Hping3*

The incident response module of the IDS can actively prevent malicious nodes from connecting to the PLC. To evaluate the attack prevention capabilities, this section performs a volumetric denial of service attack using Hping 3 [11] on the PLC from a compromised trusted node. During the attacks, this analysis varies the length of the network packets and the number of attacker threads used by Hping3. For each scenario, this section measures the time taken to respond to the attacks and the number of successful requests made by Hping3.

*a) Varying message size:* The denial of service tool Hping3 allows users to configure the size of network packets using the "-d "option. This analysis repeats the volumetric denial of service attacks ten times. During the attacks, the analysis uses the '-d' option to change the network packet sizes from 200 to 2000 bytes while keeping the number of attacker threads constant at 500. Figure 4 shows the number of allowed requests to the PLC, while Figure 5 outlines the time taken by the IDS to block the attack.

The IDS blocks DoS traffic at the same speed across payload sizes, indicating response time is independent of packet length. However, smaller packets allow more requests to slip through before the IDS can successfully block the compromised node.

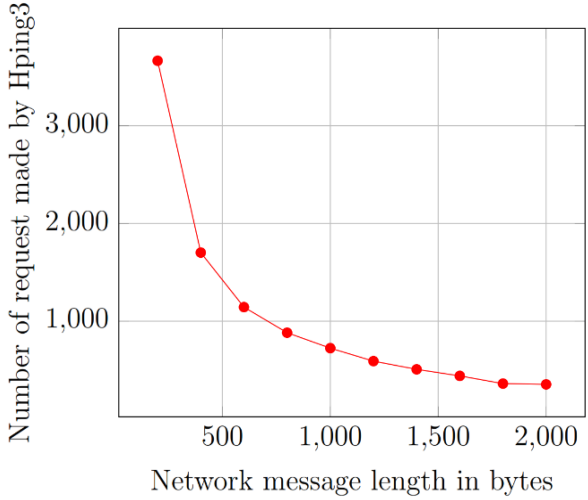

Fig. 4. Number of allowed requests admitted to the PLC under varying message sizes during volumetric DoS attack

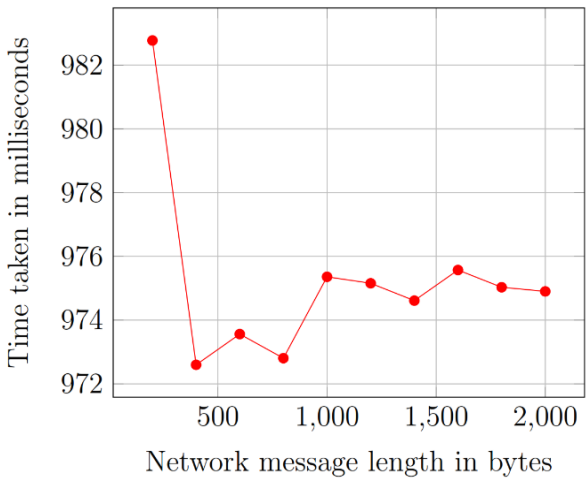

Fig. 5. Time taken by the IDS to block malicious traffic across different packet sizes in a volumetric DoS attack.

*b) Varying number of attacker threads:* This analysis repeats the volumetric denial of service attacks ten times. Blocking time does not change as we raise attacker threads from 100 to 10,000 at a fixed 1200-byte payload, as shown in Figure 6. Figure 6 shows only minor changes in the number of requests admitted before the block. The results show that increasing concurrency does not impair mitigation, and the IDS incident response system behaves workload-agnostically with respect to threading.

Figure 5 and 7 shows a stable mitigation latency that does not depend on Hping3's packet size or thread count. The incident response module blocks the source via an iptables rule and then purges residual flood traffic.

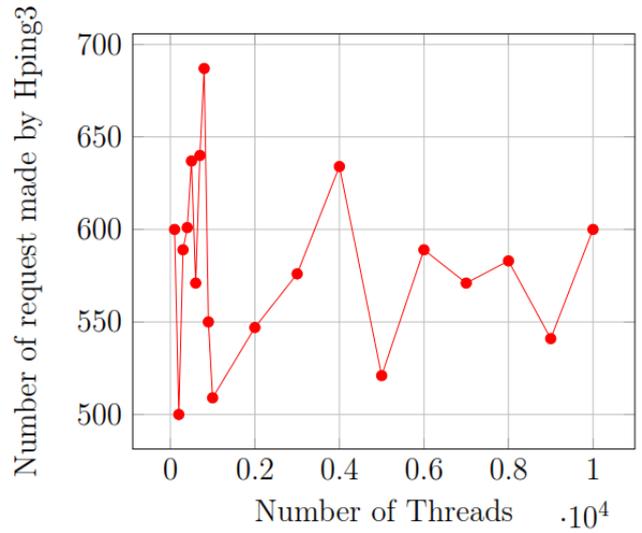

Fig. 6. Number of allowed requests to the PLC under increasing attacker thread counts during volumetric DoS attack

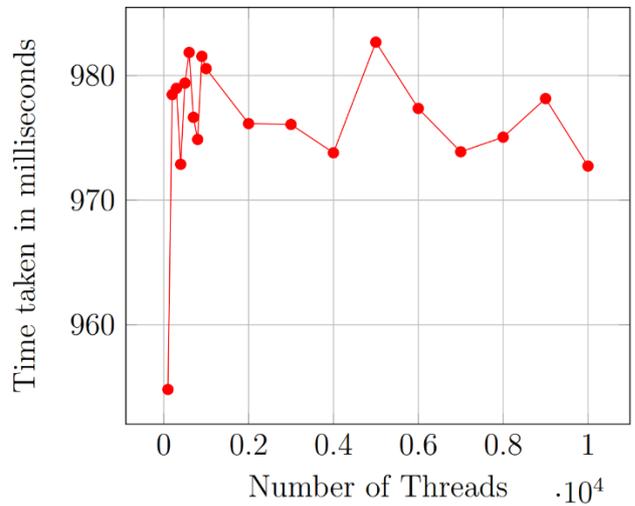

Fig. 7. Blocking time of the IDS incident response module for varying attacker thread counts during volumetric DoS attack.

## CONCLUSION AND FUTURE WORK

This research presents an embedded IDS that uses protocol-agnostic, header-level telemetry to detect network-based cyber anomalies. The IDS streams the network packets into two data pipelines. The first pipeline baselines the normal state of the network and flags deviations. When detecting an anomaly, Pipeline II applies a supervised random-forest classifier to assign an attack category. An incident response system executes a predefined set of rules to take required action against a detected cyber-attack. The embedded design adds a median of 2,031 μs of network latency and does not measurably impact the PLC's real-time scan cycle.

Future research can extend the embedded IDS beyond the midstream terminal to additional ICS domains, including electric power substations, water utilities, and networked healthcare systems. Application in wider ICS sectors can help evaluate domain-specific performance across vendors, protocols, and operating conditions. Researchers can also integrate features from the control system process and add an online autotuning process to make the IDS more resilient to changing network conditions.